\documentclass[12pt]{article}%
\usepackage{ amsmath, graphics, rotating}% 

\begin{document}
%\large
\begin{center}{\large\bf IS GRAVITY AN INTERACTION?}\end{center}

\begin{center} {\large Felix M. Lev} \end{center}
\begin{center} {\it Artwork Conversion Software Inc.,
1201 Morningside Drive, Manhattan Beach, CA 90266, USA (Email: felixlev314@gmail.com}) \end{center}

\begin{flushleft} {\bf Abstract:}
We consider a possibility that gravity is not an interaction but a 
manifestation of a symmetry based on a Galois field. \end{flushleft}

\begin{flushleft} {\bf R\'{e}sum\'{e}:} Nous consid\'{e}rons la possibilit\'{e} que la gravit\'{e} 
ne soit pas une interaction, mais une
manifestation d'une sym\'{e}trie sur la base d'un champ de Galois. \end{flushleft}

\begin{flushleft} PACS: 11.30Cp, 11.30.Ly\end{flushleft}
\begin{flushleft} Keywords: quantum theory, de Sitter invariance, gravity\end{flushleft}

\vfill\eject

The reader might immediately conclude that the title of this note is meaningless since the phenomenon
of "gravitational interaction" is well known from our everyday experience. However, lessons from
the theory of relativity and, especially, quantum theory teach us that everyday experience is not
always a best judge. 

In theoretical physics there is no unambiguous criterion for determining
whether two otherwise isolated particles interact or not. For example, in classical (i.e. non quantum) nonrelativistic
and relativistic mechanics the criterion is clear and simple: if the relative acceleration of the
particles is zero, they do not interact, otherwise they interact. However, those theories are based
on Galilei and Poincare symmetries, respectively and there is no reason to believe that those
symmetries are exact symmetries of nature.

In quantum theory, if $E$ is the energy operator of the two-body system, 
$E_1$ is the energy operator of particle 1 and $E_2$ is the energy operator of particle 2 then 
one can formally define the interaction operator $V$ such that $E=E_1+E_2+V$. Therefore the
criterion can be such that particles do not interact if $V=0$ and interact otherwise.
 
This definition can be generalized as follows. Each system is described by a set of independent 
representation operators of
the symmetry algebra $O_i$, where $i=1,2,...N$ and $N$ is the dimension of the algebra 
(for example, $N=10$ for Galilei and Poincare algebras). Let \{$O_i^{(1)}$\} be a set of operators 
describing particle 1,
\{$O_i^{(2)}$\} be a set of operators describing particle 2 and \{$O_i$\} be a set of operators 
describing the two-particle system as a whole. Then the particles do not interact if 
$O_i=O_i^{(1)}+O_i^{(2)}$ for each $i=1,2...N$ and interact otherwise. The situation when 
$O_i=O_i^{(1)}+O_i^{(2)}$ means that a representation describing a two-body system is a tensor
product of representation describing single particles. Therefore one might say that two particles
do not interact if the representation describing a two-particle system is the tensor product of the
single particle representations and interact otherwise. Such a definition is reasonable for
Galilei and Poincare symmetries but we do not know yet whether it is reasonable for other
symmetries.  

In local quantum field theory (QFT) the criterion is also clear and simple: the particles interact
if they can exchange by virtual quanta of some fields. For example, the electromagnetic interaction
between the particles means that they can exchange by virtual photons, the gravitational interaction -
that they can exchange by virtual gravitons etc. 
Although standard QFT is based on
Poincare symmetry, physicists typically believe that the notion of interaction adopted in QFT is valid for any symmetry. 

Consider now a case of de Sitter symmetries. For definiteness we will discuss the de Sitter (dS) SO(1,4) symmetry and the same
considerations can be applied to the anti de Sitter (AdS) symmetry SO(2,3). The dS spacetime is a four-dimensional
manifold in the five-dimensional space defined by $x_1^2+x_2^2+x_3^2+x_4^2-x_0^2=R^2$. Note that at a fixed value of the 
global dS time $x_0$, the dS space is a three-dimensional sphere with the radius $R_*=(R^2+x_0^2)^{1/2}$ in the four-dimensional space.
In the formal limit $R\to\infty$ the action of 
the dS group on this space proceeds to the action of the Poincare group on the Minkowski space. In the literature, instead of $R$, 
the cosmological constant $\Lambda=3/R^2$ is often used. Then $\Lambda>0$ in the dS case and $\Lambda<0$ in the AdS one. 

If one assumes that spacetime is fundamental then in the spirit of General Relativity (GR) it is natural to think that
the empty space is flat, i.e. that $\Lambda=0$. This was the subject of the well-known 
dispute between Einstein and de Sitter described in a vast literature. However, in view of the recent astronomical data,
it is now accepted that $\Lambda\neq 0$ and, although it is very small, it is rather positive than negative.
Since there exists a vast literature on dS and AdS symmetries and many authors often express considerably different opinions,
we will make a few remarks on what is important for the further presentation.  

The metric tensor on the dS space is obviously
\begin{equation}
g_{\mu\nu}=\eta_{\mu\nu}-x_{\mu}x_{\nu}/(R^2+x_{\rho}x^{\rho})
\label{1}
\end{equation}
where $\mu,\nu,\rho = 0,1,2,3$, $\eta_{\mu\nu}$ is the diagonal tensor with the components 
$\eta_{00}=-\eta_{11}=-\eta_{22}=-\eta_{33}=1$ and a summation over repeated indices is assumed. It is easy to calculate the
Christoffel symbols in the approximation where all the components of the vector $x$ are much less than $R$:
$\Gamma_{\mu,\nu\rho}=-x_{\mu}\eta_{\nu\rho}/R^2$.
Then a direct calculation shows that in the nonrelativistic approximation the equation of motion is
\begin{equation}
{\bf a}={\bf r}c^2/R^2
\label{2}
\end{equation}
where ${\bf a}$ and ${\bf r}$ are the acceleration and the radius vector of the particle, respectively.

The fact that even a single particle in the Universe has a nonzero acceleration might be treated as contradicting the law of 
inertia but this law has been postulated only for Galilei or Poincare symmetries and we have ${\bf a}=0$ in the 
limit $R\to\infty$. A more serious problem arises
%However, contradictions do arise if well known results
%from classical field theory are applied to the de Sitter symmetries. For example, it is well known that if a particle
%with the electric charge $e$ is moving with an acceleration, then the force of radiation friction acting on the particle is
%${\bf f}=(2e^2/3c^3)d{\bf a}/dt$ and therefore even for a free particle the equation of motion is $m{\bf a}=(2e^2/3c^3)d{\bf a}/dt$
%where $m$ is the particle mass.
%If the initial condition is ${\bf a}(t_0)=0$ then the solution is ${\bf a}(t)=0$ but if a free particle passed through a
%region where the electric field was not zero, its acceleration will rise as $exp(3mc^3t/2e^2)$. As noted in well known textbooks
%(see e.g. Ref. \cite{LL}), the absurdity of this result shows that classical electrodynamics is not fully self-contained. 
%However, in de Sitter theories the situation is even worse since, as follows from Eq. (4), the initial acceleration is zero only 
%if the particle is always at rest at ${\bf r}=0$.
if GR is applied for describing a free particle in the dS world. According to GR, any system moving with an acceleration 
necessarily loses energy for
emitting gravitational waves. According to the Einstein quadrupole formula, the loss of the energy is given by  
$-dE/dt=(G/45c^5)(d^3D_{ik}/dt^3)^2$ where $G$ is the gravitational constant, $D_{ik}$ is the quadrupole moment and $i,k=1,2,3$. 
For a single particle moving
along the $x$ axis, the only nonzero element of the quadrupole moment is $D_{xx}=2mx^2$ where $m$ is the particle mass. 
Therefore, as follows from Eq. (2),
$-dE/dt=4Gx^2v^2/45c^3R^2$ where $v$ is the particle velocity. We see that the loss of energy depends on the choice of the
origin in the coordinate space and one might think that this result is unphysical.

In the literature there are several different opinions on such a possibility. One might say that in the given case it is not 
legitimate to apply GR since the constant $G$ characterizes interaction between
different particles and cannot be used if only one particle exists in the world. Moreover, although GR has been confirmed in
several experiments in Solar system, it is not clear whether it can be extrapolated to cosmological distances.
More popular explanations are based on the assumption that the empty dS space is not literally empty. If the Einstein 
equations are written in the form
$G_{\mu\nu}+\Lambda g_{\mu\nu}=(8\pi G/c^4)T_{\mu\nu}$ where $T_{\mu\nu}$ is the stress-energy tensor of matter then the
case of empty space is often treated as a vacuum state of the field with the stress-energy tensor $T^{vac}_{\mu\nu}$ such that
$(8\pi G/c^4)T^{vac}_{\mu\nu}=-\Lambda g_{\mu\nu}$. This field is often called dark energy. With such an approach one implicitly
returns to Einstein's point of view that a space with $\Lambda\neq 0$ cannot be empty and treats $\Lambda\neq 0$ as a dark energy
in QFT on the flat background. 

However, in this case a new serious problem arises. Such a theory is not renormalizable and with reasonable cutoffs the quantity
$\Lambda$ in units $\hbar/2=c=1$ appears to be of order $1/l_{P}^2=1/G$ where $l_P$ is the Planck length. 
It is obvious that since in the above theory the
only dimensional quantities in units $\hbar/2=c=1$ are $G$ and $\Lambda$, and the theory does not have other parameters, the
result that $G\Lambda$ is of order unity seems to be natural. However, this value of $\Lambda$ is at least by 120 orders of magnitude
greater than the experimental one. Numerous efforts to solve this cosmological constant problem have not been successful so far although
many explanations have been proposed. We believe the situation is rather ridiculous for the following reason. If the value of $\Lambda$
were exactly zero then it would be nothing to discuss. But if $\Lambda\neq 0$ but is extremely small, one could use the same
approach and then the contradiction would be not 120 orders of magnitude but much more. In addition, many physicists argue that in the spirit
of GR, the theory should not depend on the choice of the background spacetime (so called a principle of background independence)
and there should not be a situation when the flat background is preferable.

Consider now the dS symmetry from the point of view of quantum theory. In this theory any physical quantity can be discussed 
only in conjunction with the operator defining this quantity. For example, in standard quantum mechanics  
the quantity $t$ is a parameter, which has the meaning of time only in the
classical limit since there is no operator corresponding to this quantity. The problem of how time should be defined
on quantum level is very difficult and is discussed in a vast literature. It has been also well known  
since the 1930's \cite{NW} that, when quantum mechanics is combined with 
relativity, there is no operator satisfying all the properties of the spatial position operator. In other 
words, the coordinates cannot be exactly measured even in situations when exact measurement is allowed by the
nonrelativistic uncertainty principle. In the introductory section of the well-known textbook \cite{BLP}
simple arguments are given that for a particle with mass $m$, the coordinates cannot be measured with the accuracy 
better than the Compton wave length ${\hbar}/mc$. Hence, the exact measurement is possible only
either in the nonrelativistic limit (when $c\to\infty$) or classical limit (when ${\hbar}\to 0)$.
 From the point of view of quantum theory, one can discuss if the {\it coordinates of particles} 
can be measured with a sufficient accuracy, while the notion
of empty spacetime background, regardless of whether it is flat or curved, fully contradicts basic principles 
of this theory. Indeed, the coordinates of points, which exist only in our imagination are not measurable. 

In particular, the quantity $x$ in the Lagrangian density $L(x)$ is not measurable. Note that the Lagrangian is 
only an auxiliary tool for constructing Hilbert spaces and operators and this is all we need to have the maximum 
possible information in quantum theory. After this construction has been done, one can safely forget about  
spacetime coordinates of the empty space and Lagrangian. So Lagrangian can be at best treated
as a hint for constructing a reasonable theory since a fundamental approach should not proceed from
notions, which have no meaning. As stated in Ref. \cite{BLP},
local quantum fields and Lagrangians are rudimentary notion, which will disappear in the ultimate quantum theory.
Those ideas have much in common with the Heisenberg S-matrix program
and were rather popular till the beginning of the 1970's. Although no one questioned those ideas, they
are now almost forgotten in view of successes of gauge theories.

Note that even in GR, which is a pure classical (i.e. non-quantum) theory, the meaning of reference frame is not 
quite clear. In standard textbooks (see e.g. Ref. \cite{LL}) the reference frame in GR is defined as a collection of weightless bodies, each
of which is characterized by three numbers (coordinates) and is supplied by a clock. It is
obvious that such a notion (which resembles ether) is not physical.
There is no doubt that GR is a great achievement of theoretical physics and has achieved 
great successes in describing experimental data. At the same time, it is a pure classical theory 
fully based on classical spacetime. Therefore it is unrealistic to expect that successful 
quantum theory of gravity will be based on quantization of GR. The results of GR should follow from quantum theory of gravity only 
in situations when spacetime coordinates of real bodies is a good approximation while in general the 
formulation of quantum theory might not involve spacetime at all.

If we accept that quantum theory should not proceed from empty spacetime background, a problem arises how symmetry 
should be defined on quantum level. In the spirit of Dirac's
paper \cite{Dir}, we postulate that on quantum level a symmetry means that a system is described by a set 
of operators, which satisfy certain commutation relations. 
We believe that for understanding this Dirac's idea the following example might be useful.
If we define how the energy should be measured (e.g. the energy of bound states, kinetic energy etc.),
we have a full knowledge about the Hamiltonian of our system. In particular, we know how the Hamiltonian
should commute with other operators. In standard theory the Hamiltonian is also interpreted as an
operator responsible for evolution in time, which is considered as a classical macroscopic parameter. 
In situations when this parameter is a good approximate parameter, macroscopic transformations
from the symmetry group corresponding to the evolution in time have a physical meaning. However, there is no guarantee that 
such transformations always have a physical meaning (e.g. at the very early stage of the Universe). 
In general, according to principles of quantum theory, selfadjoint operators in Hilbert spaces represent observables but
there is no requirement that parameters defining a family of unitary transformations generated by a selfadjoint operator
are eigenvalues of another selfadjoint operator. A well known example from standard quantum mechanics is that if $P_x$ is
the $x$ component of the momentum operator then the family of unitary transformations generated by $P_x$ is $exp(iP_xx/\hbar)$
where $x\in (-\infty,\infty)$ and such parameters can be identified with the spectrum of the position operator. At the same time, 
the family of unitary transformations generated by the Hamiltonian $H$ is $exp(-iHt/\hbar)$ where $t\in (-\infty,\infty)$
and those parameters cannot be identified with a spectrum of a selfadjoint operator on the Hilbert space of our system. 
In the relativistic case the parameters $x$ can be formally identified with the spectrum of the Newton-Wigner position
operator \cite{NW} but it is well known that this operator does not have all the required properties for the position
operator.

The {\it definition} of the dS symmetry on quantum level is that in units $\hbar/2=c=1$ the operators $M^{ab}$ 
($a,b=0,1,2,3,4$, $M^{ab}=-M^{ba}$) 
describing the system under consideration satisfy the commutation relations of the dS Lie algebra so(1,4), i.e. 
\begin{equation}
[M^{ab},M^{cd}]=-2i (\eta^{ac}M^{bd}+\eta^{bd}M^{ac}-
\eta^{ad}M^{bc}-\eta^{bc}M^{ad})
\label{4}
\end{equation}
where $\eta^{ab}$ is the diagonal metric tensor such that
$\eta^{00}=-\eta^{11}=-\eta^{22}=-\eta^{33}=-\eta^{44}=1$.
With such a definition of symmetry on quantum level, the dS symmetry looks more natural than the Poincare symmetry.
In the dS case all the ten representation operators of the symmetry algebra are angular momenta while in the
Poincare case only six of them are angular momenta and the remaining four operators represent standard energy and
momentum. If we define the operators $P^{\mu}$ as $P^{\mu}=M^{4\mu}/R$ then in the limit $R\to\infty$ the relations 
(3) will become the commutation relations for representation operators of the Poincare algebra such that the operators
$P^{\mu}$ are the four-momentum operators. Since in standard quantum theory the states of the system are usually
considered at fixed values of time, it is probably more physical to require that $P^{\mu}=M^{4\mu}/R_*$ but both 
definitions are equivalent when $R\gg x_0$.

A theory based on the above definition of the dS symmetry on quantum level cannot involve quantities which are 
dimensional in units $\hbar/2=c=1$. In particular, it cannot involve the dS space, the gravitational constant and 
the cosmological constant. The latter appears either as a parameter defining the relation
between the dS and Poincare symmetries in a special case when the latter is a good approximate symmetry
or if in classical limit when one wishes to work with the dS space.
Therefore with our formulation of symmetry on quantum level the cosmological constant problem does not arise at all
but instead we have a problem of why nowadays Poincare symmetry is so good approximate symmetry. This is rather a problem of 
cosmology but not quantum physics. 

By analogy with standard quantum theory, we require that, by definition, elementary particles in the dS invariant theory are described
by irreducible representations (IRs) of the dS algebra by Hermitian operators. As shown in Refs. \cite{lev1a,lev3}, such representations 
can be explitly constructed by using well known results about unitary irreducible representations (UIRs) of the dS group. 
An excellent description of such UIRs for physicists can be found in a book by Mensky \cite{Mensky}. As shown in Ref. \cite{Mensky},
they can be implemented in the Hilbert space of functions $f({\bf v})$ defined on two Lorentz hyperboloids  
$v_0=\pm (1+{\bf v}^2)^{1/2}$ such that $\int |f({\bf v})|^2d^3{\bf v}/|v_0|<\infty$. As shown in Refs. \cite{lev1a,lev3},
the action of the dS energy operator on the upper hyperboloid in the spinless case is given by
\begin{equation}
M_{04}=m_{dS} v_0+2i v_0({\bf v}\frac{\partial}{\partial {\bf v}}+\frac{3}{2})
\label{5}
\end{equation}
where $m_{dS}$ is {\it the dS mass}. 
Note that in deriving this expression and expressions for other representations operators of the so(1,4) algebra, 
no approximations have been made and the results are exact. In particular, the dS space, the cosmological constant and 
the Riemannian geometry have not been involved at all. Nevertheless, the expressions for the representation operators
is all we need to have the maximum possible information in quantum theory.

We now define $H=M_{04}/R$ and $m=m_{dS}/R$. Then in first order in $1/R$ we have that in the nonrelativistic approximation
\begin{equation}
H=m + \frac{{\bf p}^2}{2m} +\frac{2i}{R}({\bf p}\frac{\partial}{\partial {\bf p}}+\frac{3}{2})
\label{6}
\end{equation}
where ${\bf p}=m{\bf v}$. Suppose now that, by analogy with the nonrelativistic quantum mechanics, the position operator 
${\bf r}$ is meaningful and {\it can be defined} as 
$2i\partial/\partial {\bf p}$ (for a reason, which will be clear further, we accept units where $\hbar/2=1$ rather than
$\hbar=1$). Therefore, the classical nonrelativistic Hamiltonian is
$H({\bf r}, {\bf p}) ={\bf p}^2/(2m) + {\bf r}{\bf p}/R$. Then as a consequence
of the equations of motion, the relation between the acceleration and the radius-vector is again given by Eq. (2).

Our derivation of Eq. (2) shows that from the point of view 
of quantum theory, the notion of the dS space is meaningful only in the quasiclassical approximation.
In that case the cosmological constant plays a role of a macroscopic parameter, 
which shows to what extent the dS space is close to the Minkowski space. Therefore, there is no need to associate this 
constant with the dark energy or any other quantum fields. 

In quantum theory the Fock Hilbert space for a given quantum system is a tensor product of Hilbert spaces describing elementary
particles. In particular, a two-particle Hilbert space is a tensor product of the single-particle spaces. As noted above, if the
particles do not interact, then, by definition, representation operators describing a two-particle representation are sums of the corresponding
single-particle operators. So in the dS invariant theory one can use the results for IRs and calculate the mass operator of the
free two-body system. The result of calculations \cite{lev1a,lev3} is that in the approximation when
the relative distance operator can be defined with a good accuracy, the additional term in the nonrelativistic mass operator
in comparison with the Poincare theory is $V_{dS}(r) = -m_{12}r^2/(2R^2)$ where now $r$ is the relative distance and 
$m_{12}=m_1m_2/(m_1+m_2)$ is the reduced mass. As a consequence, in quasiclassical approximation the relative 
acceleration is given by the same expression (2) but now ${\bf a}$ is the relative acceleration and ${\bf r}$ is the relative
radius vector. So a standard result is again obtained without involving Riemannian geometry.    
 
The fact that two free particles in the dS world have a relative acceleration is well known for cosmologists who 
consider the dS symmetry on classical level. This effect is called the dS antigravity. 
The term antigravity in this context means that the particles repulse rather than attract each other. In the case of the 
dS antigravity the relative acceleration of two free particles is proportional (not inversely proportional!) to the distance 
between them. We have a formal contradiction with standard intuition that even if there are only two free particles in the dS 
world, their relative acceleration is not zero. By analogy with the above case of one particle in the dS world, one might
say that the reason is that the empty dS space is not literally empty and in fact it represents a third body interacting
with the particles under consideration. This is an analog of a situation when two noninteracting particles are moving in an
inhomogeneous gravitational field and then their relative acceleration is not zero. However, as noted above, attempts to treat
the dS space as a dark energy or other quantum fields on the flat background have not been successful so far.
In other words, it is probably unrealistic to think that the Poincare symmetry is fundamental while the dS one is emergent. 
We argued that from the point of view of quantum theory, the dS symmetry is more general than the Poincare one.
Then the existence of a nonzero acceleration in a free two-body system is not a real contradiction
but simply an example showing that standard physical intuition based on Galilei or Poincare symmetries
does not work in the case of the dS symmetry. 

%A reader might think that the example with the dS antigravity for two particles is pure academic since this 
%phenomenon cannot be observed experimentally. Indeed, even if $r$ has the order of the radius of the Solar system, 
%the ratio $(r/R)^2$ is extremely small. On the other hand, if $r$ has the order of cosmological distances, it
%is not possible to find an isolated two-body system such that the distance between the bodies is $r$. However,
%the goal of the above remarks is not to discuss whether the dS antigravity in the two-body system is
%realistic or not. Our main conclusion is that standard physical intuition, assuming that two particles  
%interact if and only if their relative acceleration is zero, does not work in the case of the dS symmetry. 

Since the dS symmetry is a higher symmetry with respect to the Poincare one, this example poses a problem whether gravity and possibly 
other interactions are in fact not true interactions but effective interactions arising as a result 
of interpreting a higher symmetry in terms of a lower one. In particular, a question arises whether there 
exists a higher symmetry such that, as a consequence of this symmetry, two free particles not repulse but attract 
each other such that the
relative acceleration is given by the Newtonian gravitational law and post-Newtonian corrections. 

Such a possibility is fully out of the spirit of the mainstream approach to quantum gravity. 
One of the main arguments in favor of the mainstream approach is probably as follows.
If gravitational interaction is treated as a consequence of the graviton exchange then in the nonrelativistic 
approximation the Newton gravitational law is obtained from the diagram of the one-graviton exchange in the 
same way as the Coulomb law is obtained from the diagram of the one-photon exchange. Also the results on binary 
pulsars (the present status of this problem can be found e.g. in Ref. \cite{WT} and references therein) are
treated as a strong indirect indication of the existence of gravitons. In reality only some electromagnetic 
radiation is detected. Then one describes this radiation assuming that it belongs to a binary pulsar. 
One constructs models where the masses of the pulsars, their distances from each other etc. are adjustable 
parameters and it is assumed that the interaction of the pulsars with the interstellar matter
is weak. Then by fitting the parameters, the Einstein quadrupole formula is reproduced with a high accuracy. 

We believe that although those arguments are very serious, they should not be accepted without any reservations. 
For example, the analysis performed in Ref. \cite{WT} is based only on classical GR and 
quantum effects are not considered. Therefore, even if the data are described by the Einstein quadrupole formula, 
the fact that on quantum level the main contribution to the standard nonrelativistic gravity is the graviton exchange, is an 
additional assumption based on belief that gravity can be described in QFT or its generalizations
(string theory, loop quantum gravity etc.). One might wonder why this belief is so strong in spite of the fact 
that numerous efforts to construct quantum theory of gravity without infinities has not been successful yet
(see also the discussion below).

There exist several approaches where gravity is not fundamental but emergent. 
A very recent example is the discussion of the approach to gravity proposed by Verlinde. In Ref. \cite{Verlinde} and references therein the 
reader can find a detailed discussion of this and other approaches.  

If we wish to investigate the above idea that gravity is simply a manifestation of a higher symmetry then 
we should investigate a) what a possible candidate for this symmetry is; b) how this symmetry gives the Newton
gravitational law, post-Newtonian corrections etc. This investigation requires extensive calculations which
have been discussed in Ref. \cite{lev3} and our subsequent papers. In genre of essay it is not possible to reproduce 
those calculations and in the remaining part of this paper we will sketch only basic ideas.

For majority of physicists the fact that standard theory describes many experimental data with an unprecedented accuracy
is much more important than the fact the mathematical substantiation of the theory is rather poor.
As a consequence, the issue of infinities is probably the most challenging problem in 
standard formulation of quantum theory. 

Mathematical problems of quantum theory are discussed in a vast literature. For example, in the
well known textbook \cite{Bogolubov} it is explained in details that interacting quantized fields can only be treated 
as operatorial distributions and hence their product at the same point is not well defined.
One of ideas of the string theory is that if a point (a zero-dimensional object) is replaced by a string
(a one-dimensional object) then there is hope that infinities will be less singular. 

There exists a wide literature aiming to solve the difficulties of the theory by replacing the field
of complex numbers by quaternions, p-adic numbers or other constructions. A modern state-of-the-art of the p-adic
theory can be found, for example, in Ref. \cite{Dragovich}. At present it is not clear how to overcome all the 
difficulties but at least from the point of view of the problem of infinities a natural approach is to consider 
quantum theory over Galois fields (GFQT). Since any Galois field is finite, the 
problem of infinities in GFQT does not exist in principle and all operators are well defined. The idea of using
finite fields in quantum theory has been discussed by several authors (see e.g. Refs. \cite{Galois,Volovich}). 
As stated in Ref. \cite{Volovich}, a fundamental theory can be based either on p-adic numbers or finite fields.
In that case, a correspondence with the standard theory will take place if the number $p$ in the p-adic theory or
as a characteristic of a finite field is rather large.

The authors of Ref. \cite{Volovich} and many other papers argue that fundamental quantum theory cannot 
be based on mathematics using standard geometrical objects (such as strings etc.) at Planck distances.  
We believe it is rather obvious that the notions of continuity, differentiability, smooth manifolds etc. are 
based on our macroscopic experience. For example, the water in the ocean can be described by
equations of hydrodynamics but we know that this is only an approximation since matter 
is discrete. Therefore continuous geometry probably does not describe physics even at distances much greater than
the Planck length.   

In our opinion, an approach based on finite fields is very attractive  for 
solving problems in quantum theory as well as for philosophical and aesthetical reasons. 
Below we describe some arguments in favor of this opinion. 

The key ingredient of standard mathematics is the notions of
infinitely small and infinitely large numbers. The notion of infinitely small numbers is based on our
everyday experience that any macroscopic object can be divided by 2, 10, 1000 etc. In the view of the
existence of elementary particles, the notion of division has a limited applicability. For example, we 
cannot divide the electron or neutrino by two. Therefore, {\it if we accept the existence of elementary
particles, we should acknowledge that our experience based on standard mathematics is not universal}.

The notion of infinitely large numbers is based on the belief that {\it in principle} we can operate with any
large numbers. In standard mathematics this belief is formalized in terms of axioms (accepted without proof) 
about infinite sets (e.g. Zorn's lemma or Zermelo's axiom of choice). At the same time, in the spirit of 
quantum theory, there should be no statements accepted without proof since only those statements have physical significance, 
which can be experimentally verified, at least in principle.

For example, we cannot verify that $a+b=b+a$ for any numbers $a$ and $b$. 
Suppose we wish to verify that 100+200=200+100. In the spirit of quantum theory, it is insufficient to say
that 100+200=300 and 200+100=300. To check these relationships, we should describe an experiment where they
can be verified. In particular, we should specify whether we have enough resources to represent the numbers 100, 200 and 300. 
We believe the following observation is very important: although standard mathematics is a part of our everyday life, people typically
do not realize that {\it standard mathematics is implicitly based on the assumption that one can have any desirable 
amount of resources}.

Suppose that our Universe is finite. This implies that the amount of resources cannot be infinite and
it is natural to assume that there exists a number $p$ such that all calculations can
be performed only modulo $p$. In this case, one might consider a quantum theory over a Galois field with
the characteristic $p$. Since any Galois field is finite, the fact that arithmetic in this field is correct can be
verified, at least in principle, by using a finite amount of resources.

If one accepts the idea to replace complex numbers by a Galois field, the problem arises what
formulation of the standard quantum theory is most convenient for that purpose. A well known 
historical fact is that originally quantum theory has been proposed in two formalisms
which seemed essentially different: the Schroedinger wave formalism and the Heisenberg
operator (matrix) formalism. It has been shown later by Born, von Neumann and others that both
formalisms are equivalent and, in addition, the path integral formalism has been developed. 

In the spirit of the wave or path integral approach one might try to replace classical spacetime 
by a finite lattice which may even not be  a field. In that case the problem arises what  the natural
'quantum of spacetime' is and some of physical quantities should necessarily have the field structure. A 
detailed discussion can be found in Ref. \cite{Galois} and references therein. In contrast
to these approaches, we propose to generalize the standard operator formulation, where quantum 
systems are described by elements of a projective complex Hilbert spaces and physical quantities
are represented by selfadjoint operators in such spaces. 
For this reason GFQT could be defined as a theory where
\begin{itemize}
\item {\it Quantum states are represented by elements of a linear projective space over a Galois field 
and physical quantities are represented by linear operators in that space.}
\end{itemize}

As noted in Ref. \cite{Dragovich} and references therein, in the p-adic theory a problem arises what
number fields (if any) are preferable and there should be quantum fluctuations not only of metrics and 
geometry but also of the number field. Volovich \cite{Volovich} proposed the following number field invariance principle:
fundamental physical laws should be invariant under the change of the number field. Analogous questions 
can be posed in GFQT.

It is well known (see e.g. standard textbooks \cite{VDW}) that any Galois field can contain only $p^n$ elements where $p$
is prime and $n$ is natural. Moreover, the numbers $p$ and $n$ define the Galois field up to isomorphism. 
It is natural to require that there should exist a correspondence between any new theory and the old one, i.e. at 
some conditions both theories should give close predictions. In particular, there should exist a large number of 
quantum states for which the probabilistic interpretation is valid. Then, as shown in  
our papers \cite{lev3,lev2}, in agreement with Refs. \cite{Galois,Volovich}, the number $p$ should 
be very large. Hence, we
have to understand whether there exist deep reasons for choosing a particular value of $p$ or it is simply an accident
that our Universe has been created with this value. Since we don't know the answer, we accept a simplest version of
GFQT, where there exists only one Galois field with the characteristic $p$, which is a universal constant for
our Universe. Then the problem arises what the value of $n$ is. Since there should exist a correspondence between GFQT and the 
complex version of standard quantum theory, a natural idea is to accept that the principal
number field in GFQT is the Galois field analog of complex numbers which is constructed below.

Let $F_p=Z/pZ$ be the residue field modulo $p$ and $F_{p^2}$ be a set of $p^2$ elements $a+bi$ 
where $a,b\in F_p$ and $i$ is a formal element such that $i^2=-1$. The question arises whether $F_{p^2}$ is a
field, i.e. one can define all the arithmetic operations except division by zero.
The definition of addition, subtraction and multiplication in $F_{p^2}$
is obvious and, by analogy with the field of complex numbers, one could define division as
$1/(a+bi)\,=a/(a^2+b^2)\,-ib/(a^2+b^2)$ if $a$ and $b$ are not equal to zero simultaneously.
This definition can be meaningful only if $a^2+b^2\neq 0$ in $F_p$. If $a$ and $b$ are not
simultaneously equal to zero, this condition can obviously be reformulated such that $-1$ should not
be a square in $F_p$ (or in terminology of number theory it should not be a quadratic residue).
We will not consider the case $p=2$ and then $p$ is necessarily odd.
Then we have two possibilities: the value of $p\,(mod \,4)$ is either 1 or 3. The well known result of number theory
is that -1 is a quadratic residue only in the former case and a quadratic nonresidue
in the latter one, which implies that the above construction of the field $F_{p^2}$ is correct only if
$p=3\,\,(mod \,4)$.

The main idea of establishing the correspondence between
GFQT and the standard theory is as follows (see Refs. \cite{lev3,lev2} for a detailed discussion). The first step is to
notice that the
elements of $F_p$ can be written not only as $0,1,...p-1$ but also as $0,\pm 1,...,\pm (p-1)/2$. Such elements of $F_p$ are called minimal
residues \cite{VDW}. Since the field $F_p$ is cyclic, it is convenient to visually depict
its elements by the points of a circumference of the radius $p/2\pi$ on the plane $(x,y)$ such that  
the distance between neighboring elements of the field is equal to unity, and the elements
0, 1, 2,... are situated on the circumference counterclockwise. At the same time we depict the elements of $Z$ as usual,
such that each element $z\in Z$ is depicted by a point with the coordinates $(z,0)$.
In Fig. 1 a part of the circumference near the origin is depicted.
\begin{figure}[!ht]
\centerline{\scalebox{1.1}{\includegraphics{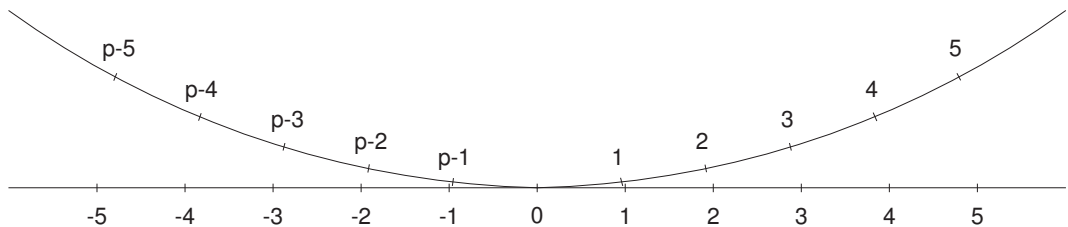}}}
\caption{
  Relation between $F_p$ and the ring of integers
}
\label{Fig.1}
\end{figure} 
Let $f$ be a map from $F_p$ to $Z$ such that $f(a)$ has the same notation in $Z$ as its minimal residue in
$F_p$. Then for elements $a,b\in F_p$ such that $|f(a)|,|f(b)|\ll p$, addition, subtraction 
and multiplication in $F_p$ and $Z$ are the same, i.e. $f(a\pm b)=f(a)\pm f(b)$ and $f(ab)=f(a)f(b)$. 

The second step is to establish a correspondence between Hilbert spaces in standard theory and spaces over a
Galois field in GFQT. We first note that Hilbert spaces contain a big redundancy of elements and we do not need to know all of
them. With any desired accuracy we can approximate each element $\tilde x$ from a Hilbert space $H$ by a finite linear combination
$ \tilde x =\tilde c_1 \tilde e_1+\tilde c_2 \tilde e_2+...\tilde c_n\tilde e_n $
where $(\tilde c_1,\tilde c_2,...\tilde c_n)$ are rational complex numbers. 
In turn, the set of such elements is redundant too. We can use the fact that Hilbert
spaces in quantum theory are projective: $\psi$ and $c\psi$ represent the same physical state, which 
reflects the fact that not the probability itself but the relative probabilities of different measurement outcomes
have a physical meaning. Therefore we can
multiply both parts of the above equality by a common denominator of the numbers
$(\tilde c_1,\tilde c_2,...\tilde c_n)$. As a result, we can always assume that
$\tilde c_j=\tilde a_j +i\tilde b_j$ where $\tilde a_j$ and $\tilde b_j$ are integers. 

Consider now a space over $F_{p^2}$ and let $x =c_1 e_1+c_2 e_2+...c_n e_n$ be a decomposition of a state $x$ over a basis
$(e_1,e_2...)$ in this space. We can formally define a scalar product $(e_j,e_k)$ such that $f((e_j,e_k))=(\tilde e_j,\tilde e_k)$.
Then the correspondence between the states $x$ and ${\tilde x}$ can be defined such that
$c_j=a_j+ib_j$ $(j=1,2...)$, $f(a_j)=\tilde a_j$ and $f(b_j)=\tilde b_j$ . If the numbers in question are much less than $p$ 
then the standard description and that based on GFQT give close
experimental predictions. At the same time, in GFQT a probabilistic interpretation is not universal and is valid only when the numbers
in question are much less than $p$.

The above discussion has a well known historical analogy. For many years people believed
that our Earth was flat and infinite, and only after a long period of time they realized that
it was finite and had a curvature. It is difficult to notice the curvature when we deal only with
distances much less than the radius of the Earth. Analogously one might think that
the set of numbers describing physics has a curvature defined by a very large number $p$ but we do not notice
it when we deal only with numbers much less than $p$. 

One might wonder that if physical quantities can take only values less than $p$ then why the values $p, p+1, p+2$ etc.
are not possible. Also one might think that if dS energies, momenta or other
quantities are close to $p$ then such a situation will have a catastrophic effect on 
physical predictions. A historical analogy is that when special relativity predicted that the speed of any particle
cannot be infinite and is limited above by the value of $c$, classical physicists treated this fact as unphysical
since from the point of view of classical physics (and everyday experience), if $v<c$ is possible then it is not clear why 
$v=1.1c$ or even $v=10c$ is not. The answer is that we cannot extrapolate our experience at $v\ll c$ to
situations when $v$ is of order $c$. In the case of GFQT, the values of physical quantities, which are greater or equal than $p$
cannot exist since this contradicts arithmetic over a Galois field. Since $p$ is probably an extremely huge number (see below), our 
experience cannot tell us what happens if some physical quantities are close to $p$, this is {\it terra incognita}. 
One might hope that further investigations will shed light on this problem. 

Since we treat GFQT as a more general theory than the standard one, it is desirable not to postulate
that GFQT is based on $F_{p^2}$ (with $p=3\,\,(mod \,4)$) because standard theory is based on complex numbers
but vice versa, explain the fact that standard theory is based on complex numbers since GFQT is based 
on $F_{p^2}$. Hence, one should find a motivation for the choice of $F_{p^2}$ in GFQT. A possible motivation 
is discussed in Refs. \cite{lev2,complex}. 

The above discussion shows that GFQT is a more general theory than the standard one since the latter is a special case
of the former when $p\to\infty$. In the approximation when $p$ is very large, GFQT can reproduce all the standard
results of quantum theory. At the same time, GFQT is well defined mathematically since it does not contain
infinities. Therefore, one can replace the standard dS or AdS symmetries by ones based on dS or AdS algebras
but not over real numbers but over a Galois field. Note that while in standard theory the dS and AdS algebras are
"better" than the Poincare algebra from aesthetic considerations (see the discussion above), in GFQT the
Poincare algebra over a Galois field is probably unphysical (see the discussion in Refs. \cite{lev3,lev2}). 

Now the dS symmetry means that the operators $M^{ab}$
satisfy the same commutation relations as above but those operators are considered in spaces over a Galois field.
Such operators implicitly depend on $p$ but they still do not depend on $R$ or $\Lambda$. Note that GFQT cannot contain
quantities dependent on units of measurements since in Galois fields there are no fractions. GFQT can contain only quantities, which 
in units $\hbar/2=c=1$ are integers. In particular, in this units the spins of fermions are odd and the spins of bosons 
are even. Note also that even dS masses of elementary particles are very large. For example, if $R$ is of order $10^{28}cm$ then
the dS mass of the electron is of order $m_eR\approx 10^{39}$, where $m_e$ is the standard electron mass. It is also obvious that
GFQT cannot contain the gravitational constant G. Then in view of our proposal, a question arises how in GFQT one can obtain the
Newton gravitational law, post - Newtonian corrections etc. This problem requires numerous complicated calculations
which now are underway. For this reason we will describe only main ideas.  

Since the phenomenon of gravity has been observed only on classical level, one needs to understand what GFQT can predict if
quasiclassical approximation is valid. Let us first recall basic properties of quasiclassical approximation in standard
quantum mechanics. In the one-dimensional case the quasiclassical wave function in the coordinate representation has the
form $\psi(x)=a(x)exp(ikx)$ where the amplitude $a(x)$ has a sharp maximum near $x=x_0\in [x_1,x_2]$ and $\Delta x=x_2-x_1\ll |x_0|$ 
is the width of the maximum. The last inequality is a requirement that the quasiclassical coordinate should be much greater
than its uncertainty.
The value of $k$ can be identified with the quasiclassical momentum of the particle if $|ka(x)|\gg |da(x)/dx|$. Indeed,
since the momentum operator is $-id/dx$, this condition ensures that the wave function is approximately the eigen function
of the momentum operator with the eigenvalue $k$. Since $|da(x)/dx|$ is of order $|a(x)/\Delta x|$, we have a condition
$k\Delta x \gg 1$. Since the uncertainty of momentum is $1/\Delta x$, this implies that the quasiclassical momentum should be
much greater than this uncertainty. At the same time, $k\Delta x$ is approximately the number of oscillations which the exponent 
makes on the segment $[x_1,x_2]$. Therefore the number of oscillations should be much greater than unity. In particular, the
quasiclassical approximation cannot be valid if $\Delta x$ is very small, but on the other hand, $\Delta x$ cannot be
very large since it should be much less than $x_0$.    

As already noted, the fact that there exists a maximum number $p$ means that the probabilistic interpretation of quantum
theory can be valid only if parameters defining the wave function are much less than $p$ \cite{lev3,Galois}.
In that case the parameters can be treated as complex integers. The wave function of a body is a product of the wave functions 
of the constituents. Since the magnitude of the product of 
nonzero integers is usually greater than each multiplier, 
the parameters of the $N$-particle wave function are closer to $p$ if $N$ increases and therefore the mass of the
body becomes greater. 

Each body is comprised of nucleons and electrons. It is reasonable to assume that the width of the 
nucleon momentum wave functions in the body is much greater than the width of the electron wave functions. Indeed, the width
of the electron wave functions is of the order of $1/l_B$, where $l_B$ is the Bohr radius while the width of the nucleon wave
functions is of order of $1/l_{\pi}$ where $l_{\pi}$ is the pion Compton wave length. Suppose that a body with the mass $m_1$ 
contains $N_1$ nucleons. Then
a reasonable model for the momentum wave function of the macroscopic body is such that the width of the momentum distribution
cannot be greater than $lnp/(N_1R)$ since otherwise the probabilistic interpretation will be broken.
Let $k$ be the relative two-body momentum and $r$ be the relative distance. Then there exist models of the two-body wave function
where the width is of order $\Delta k\approx lnp /((N_1+N_2)R)$ where $m_2$ is the mass of the second body and $N_2$ is the
number of nucleons in that body. By analogy with the above example in quantum mechanics, the
condition that the state should be quasiclassical is $r\Delta k\gg 1$. Therefore the parameter defining a correction to the
quasiclassical approximation is of order $1/(r\Delta k)$. It is easy to show that this correction gives a negative contribution
to the two-body energy and in the nonrelativistic approximation this contribution equals $-Gm_1m_2/r$ where $G$ is of order 
$Rl_N/lnp$ and $l_N$ is the Compton wave length of the nucleon. 

Note that in this scenario $G$ depends on the model of the wave function of the macroscopic body, on $p$ and on $R$.
The latter dependence is a consequence of the fact that we calculate the correction not to the dS energy but to the Poincare
energy in the approximation when Poincare invariance is a good approximate symmetry. We see that if $p$ is large but not infinitely
large, the width of the momentum distribution cannot be such that the quasiclassical approximation is valid with any desired
accuracy. Since $G=l_P^2$, the value of $lnp$ is of order $Rl_N/l_P^2$. Therefore if $R$ is order $10^{28}cm$ then,
taking into account that $l_P$ is of order $10^{-33}cm$ and $l_N$ is of order $10^{-14}cm$, we get that $lnp$ is of order
$10^{80}$ and $p$ is of order $e^{10^{80}}$. Therefore, $p$ is indeed an extremely huge number. In such a scenario it is 
clear that gravity necessarily requires that at least one body should be macroscopic. In particular, in this picture gravity 
does not exist for systems where all the bodies are elementary particles. The details of calculations will be described elsewhere. 

In the beginning of this essay we noted that in standard approach to the cosmological constant problem the conclusion that $G\Lambda$ is
of order unity seems to be natural since the theory does not contain other parameters. However, in GFQT we have an additional parameter
$p$. Then in the above scenario the value of $G$ is much less than $1/\Lambda$ since $G$ depends on $p$ as $1/lnp$. 

We see that in contrast with modern approaches to gravity, $G$ is not a fundamental constant. In this connection
note that as pointed out in Ref. \cite{Uzan}, {\it "Contrary to most of the other fundamental constants, 
as the precision of the measurements increased, the disparity between the measured values of $G$ also
increased. This led the CODATA in 1998 to raise the relative uncertainty for $G$ from
0.013\% to 0.15\%".} Several well known physicists, including Dirac, discussed a possibility
that $G$ is not fundamental and, in particular, it is time dependent. In the above scenario $p$ is a fundamental constant but $R$
is not. We have also noted, that $R_*$ is probably more physical than $R$ as a parameter characterizing the accuracy of the
Poincare symmetry. Since the dimension of $G$ is $length^2$, it is reasonable that $G$ becomes greater when $R_*$ is greater.
Since at present the Universe is expanding, one might expect that $R_*$ is increasing and therefore $G$ is increasing too.

Note added in proof: a more detailed discussion of the cosmological constant problem can be found in arXiv:1004.1861.

{\it Acknowledgements: } I am grateful to Volodya Netchitailo for stimulating discussions and to the anonymous referee of this
essay for important remarks.

\end{document}